\documentclass[aps,twocolumn,superscriptaddress,showpacs]{revtex4-1}

\usepackage{graphicx}
\usepackage{subfigure}
\usepackage{hyperref}
\usepackage{units}
\usepackage{amsmath}

\begin{document}

% Use the \preprint command to place your local institutional report
% number in the upper righthand corner of the title page in preprint mode.
% Multiple \preprint commands are allowed.
% Use the 'preprintnumbers' class option to override journal defaults
% to display numbers if necessary
%\preprint{}

%Title of paper
\title{Resistive properties and phase diagram of the organic antiferromagnetic metal $\kappa$-(BETS)$_2$FeCl$_4$}

\author{Michael Kunz}
%\email[]{Your e-mail address}
%\homepage[]{Your web page}
%\thanks{}
%\altaffiliation{Physik-Department, Technische Universit\"{a}t M\"{u}nchen, Garching, Germany}
\affiliation{Walther-Mei\ss{}ner-Institut, Bayerische Akademie der Wissenschaften, Garching, Germany}
\affiliation{Physik-Department, Technische Universit\"{a}t M\"{u}nchen, Garching, Germany}

\author{Werner Biberacher}
%\email[]{Your e-mail address}
%\homepage[]{Your web page}
%\thanks{}
%\altaffiliation{}
\affiliation{Walther-Mei\ss{}ner-Institut, Bayerische Akademie der Wissenschaften, Garching, Germany}

\author{Natalia D. Kushch}
%\email[]{Your e-mail address}
%\homepage[]{Your web page}
%\thanks{}
%\altaffiliation{}
\affiliation{Institute of Problems of Chemical Physics, Chernogolovka, Russia}

\author{Akira Miyazaki}
\affiliation{%Department of Environmental Applied Chemistry,
      Faculty of Engineering,
      University of Toyama,
%      3190 Gofuku, Toyama-shi
      Toyama,
      %930-8555
      Japan
}
\author{Mark V. Kartsovnik}
\email[]{mark.kartsovnik@wmi.badw.de}
%\homepage[]{Your web page}
%\thanks{}
%\altaffiliation{}
\affiliation{Walther-Mei\ss{}ner-Institut, Bayerische Akademie der Wissenschaften, Garching, Germany}

\date{\today}

\begin{abstract}
The low-temperature electronic state of the layered organic charge-transfer salt $\kappa$-(BETS)$_2$FeCl$_4$ was probed by interlayer electrical resistance measurements under magnetic field.
Both above and below $T_{\mathrm{N}}=0.47$\,K, the temperature of antiferromagnetic ordering of $3d$-electron spins of Fe$^{3+}$ localized in the insulating anion layers, a non-saturating linear $R(T)$ dependence has been observed. A weak superconducting signal has been detected in the antiferromagnetic state, at temperatures $\leq 0.2$\,K. Despite the very high crystal quality, only a tiny fraction of the sample appears to be superconducting. Besides a small kink feature in the resistivity, the impact of the antiferromagnetic ordering of localized Fe$^{3+}$ spins on the conduction $\pi$-electron system is clearly manifested in the Fermi surface reconstruction, as evidenced by Shubnikov-de Haas oscillations. The "magnetic field -- temperature" phase diagrams for the field directions parallel to each of the three principal crystal axes have been determined. For magnetic field along the easy axis a spin-flop transition has been found. Similarities and differences between the present material and the sister compound $\kappa$-(BETS)$_2$FeBr$_4$ are discussed.
\end{abstract}

% insert suggested PACS numbers in braces on next line
% insert suggested keywords - APS authors don't need to do this
%\keywords{}

%\maketitle must follow title, authors, abstract, \pacs, and \keywords
\maketitle

\section{Introduction}

Bifunctional organic charge transfer salts combining nontrivial electrical conduction and magnetic properties gained significant attention in the past two decades. Of special interest have been the isomeric families $\lambda$- and $\kappa$-(BETS)$_2$FeX$_4$, where BETS stands for bis(ethylenedithio)tetraselenafulvalene and X\,= Cl or Br \cite{kobayashi_organic_2004}. The members of this family can be considered as natural nanoscale heterostructures in which two-dimensional (2D) conducting layers of organic BETS donors (the Greek symbols $\lambda$ and $\kappa$ denote different packing motifs of BETS molecules \cite{ishi98}) are alternated with magnetic insulating layers of anions FeX$_4^-$. The electronic ground states of these compounds are determined by the interplay of two main factors. On the one hand, electronic correlations in the relatively narrow \nicefrac{1}{2}-filled conduction band formed by $\pi$-electrons in the BETS layers give rise to the insulating Mott instability. On the other hand, there is significant exchange interaction between localized $d$-electron spins of neighboring Fe$^{3+}$ ions in the insulating anion layers as well as between the localized $d$-electrons of Fe$^{3+}$ and itinerant $\pi$-electrons of BETS layers. A spectacular manifestation of the role of magnetic interactions in forming the electronic ground state has been found in $\lambda$-(BETS)$_2$FeCl$_4$.
At low temperatures this salt is in an insulating state with both the $d$- and $\pi$-electron spin subsystems being antiferromagnetically ordered \cite{kobayashi_new_1993,goze_magnetotransport_1994,brossard_interplay_1998}. Application of a magnetic field first leads to a re-entrance to the paramagnetic (PM) metallic state at about 10\,T \cite{brossard_interplay_1998} and -- most exciting -- to a field-induced superconducting (SC) state at even higher fields, starting from 17\,T \cite{uji_magnetic-field-induced_2001}. The latter is caused by the Jaccarino-Peter compensation effect \cite{jaccarino_ultra-high-field_1962}, which is a direct consequence of the $\pi$-$d$ exchange.

The isomer salt $\kappa$-(BETS)$_2$FeCl$_4$ (hereafter referred to as $\kappa$-Cl) and its sister compound $\kappa$-(BETS)$_2$FeBr$_4$ ($\kappa$-Br) also show antiferromagnetic (AFM) ordering in the Fe$^{3+}$ $d$-electron spin subsystem \cite{kobayashi_new_1993,kobayashi_new_1996}. However, by contrast to the above-mentioned $\lambda$-phase salt, they were found to stay metallic and even become superconducting in the AFM state \cite{konoike_magnetic-field-induced_2004,otsuka_organic_2001}.
A likely reason for that is weaker magnetic interactions. Indeed, the N\'{e}el temperatures here are considerably lower: $T_{\mathrm{N}} = 0.45$\,K and $2.4$\,K, for $\kappa$-Cl \cite{otsuka_organic_2001} and $\kappa$-Br, \cite{ojima_antiferromagnetic_1999} respectively. For the latter salt the $\pi$-$d$ exchange field has been estimated as $\approx 12$\,T \cite{mori_estimation_2002,cepas_magnetic-field-induced_2002}, which is about 3 times lower than for $\lambda$-(BETS)$_2$FeCl$_4$ \cite{balicas_superconductivity_2001,mori_estimation_2002,cepas_magnetic-field-induced_2002}. Although this exchange field is not strong enough for driving the conduction system into the spin-ordered insulating state, it is clearly manifested in the Fermi surface reconstruction \cite{konoike_fermi_2005} and in the field-induced SC transition  \cite{fujiwara_indication_2002,konoike_magnetic-field-induced_2004}.

For the $\kappa$-Cl salt the available experimental data is much more limited, probably due to considerably lower characteristic temperatures $T_{\mathrm{N}}$ and $T_c$.
As mentioned above, $\kappa$-Cl is also reported to be an AFM superconductor with $T_c \sim 0.1$\,K according to a.c.-susceptibility measurements \cite{otsuka_organic_2001}. Muon spin rotation experiment \cite{pratt_sr_2003} has also revealed an anomaly between $0.1$ and $0.2$\,K, which was attributed to a SC transition. However, no other experiments, e.g. transport or specific heat have been able to detect superconductivity in this compound as yet \cite{otsuka_organic_2001}. The $\pi$-$d$ exchange field is predicted to be even lower than for $\kappa$-Br \cite{mori_estimation_2002}.
%Theoretically it was estimated as $\mu_0H_e = 5.8$\,T.
However, its experimental evaluation is still lacking.

While the magnetization anisotropy seems to be characterized by the same principal axes as in $\kappa$-Br \cite{otsuka_organic_2001}, no detailed study of the influence of the magnetic field on the electronic properties and ground state has been done, to the best of our knowledge.

In order to gain more information on the interplay between the conducting and magnetic subsystems of the $\kappa$-Cl salt and compare it with the other compounds of the (BETS)$_2$FeX$_4$ family, we have carried out measurements of its interlayer resistance at temperatures down to $22$\,mK. While no signature of a bulk SC transition has been found, a weak deviation of the resistance from the otherwise monotonic linear $R(T)$ dependence detected at $T < 0.21$\,K is most likely caused by superconductivity arising in a minor part of the sample. The $\pi$-$d$ coupling is manifested in  a clear resistive anomaly accompanying the AFM transition both in the temperature and in the magnetic field sweeps. We have used this anomaly for delineating the magnetic phase diagram of $\kappa$-Cl for fields along all three principal axes of magnetization. We also report on Shubnikov-de Haas (SdH) oscillations in the AFM state. Similarly to the case of the $\kappa$-Br salt, the oscillations clearly point to a Fermi surface reconstruction caused by the AFM ordering.

\section{Experimental \label{sec:experimental}}

Crystals of $\kappa$-(BETS)$_2$FeCl$_4$ were synthesized by an electrocrystallization procedure as described by Kobayashi \textit{et al.} \cite{kobayashi_new_1996}. The crystals grew as regular rhombic plates with lengths of the diagonals up to 0.8\,mm in the plane of the conducting layers (crystallographic $ac$-plane) and a thickness of $\leq0.1$\,mm. The longer diagonal was along the crystallographic $a$-axis, the shorter along the $c$-axis. For measuring the interlayer (parallel to the $b$-axis) resistance, annealed 20\,$\mu$m thick platinum wires were attached to the samples in the conventional 4-probe configuration by carbon paste. Contact resistances below $15\,\Omega$ were achieved. The samples were then mounted at the cold finger of a home-built dilution refrigerator, which provided a cooling power of $5.1\,\mu$W at $T=100$\,mK and end temperatures of about 20\,mK.
Due to a considerable temperature dependence of the sample resistance till the lowest temperatures (see Section~\ref{sec:R(T)}), a heating effect of the measurement current $I_s$ could be properly evaluated. At low temperatures different values of the current were applied, ranging from 0.2 to 10\,$\mathrm{\mu A}$. A comparison of the $R(T)$ curves for the different currents showed that the sample was overheated to 180\,mK at $I_s=10\,\mathrm{\mu A}$, at the base temperature of the refrigerator. For $I_s=1\,\mathrm{\mu A}$ already no overheating above 40\,mK was detected. Zero-field measurements were conducted at $I_s=1\,\mathrm{\mu A}$. For a typical sample resistance of $20$\,m$\Omega$ the relative noise level was $\sim 2\%$.
For measurements in magnetic fields this accuracy was not sufficient to clearly resolve the transition anomalies. Therefore, the measurements of the phase diagram and of the SdH effect were done at $I_s=10\,\mathrm{\mu A}$. Thus, for those measurements the lowest temperature was limited to 180\,mK and an appropriate correction for the overheating was made.

\begin{figure}[tb]
	\centering
		\subfigure[]{\includegraphics[width=0.14\textwidth]{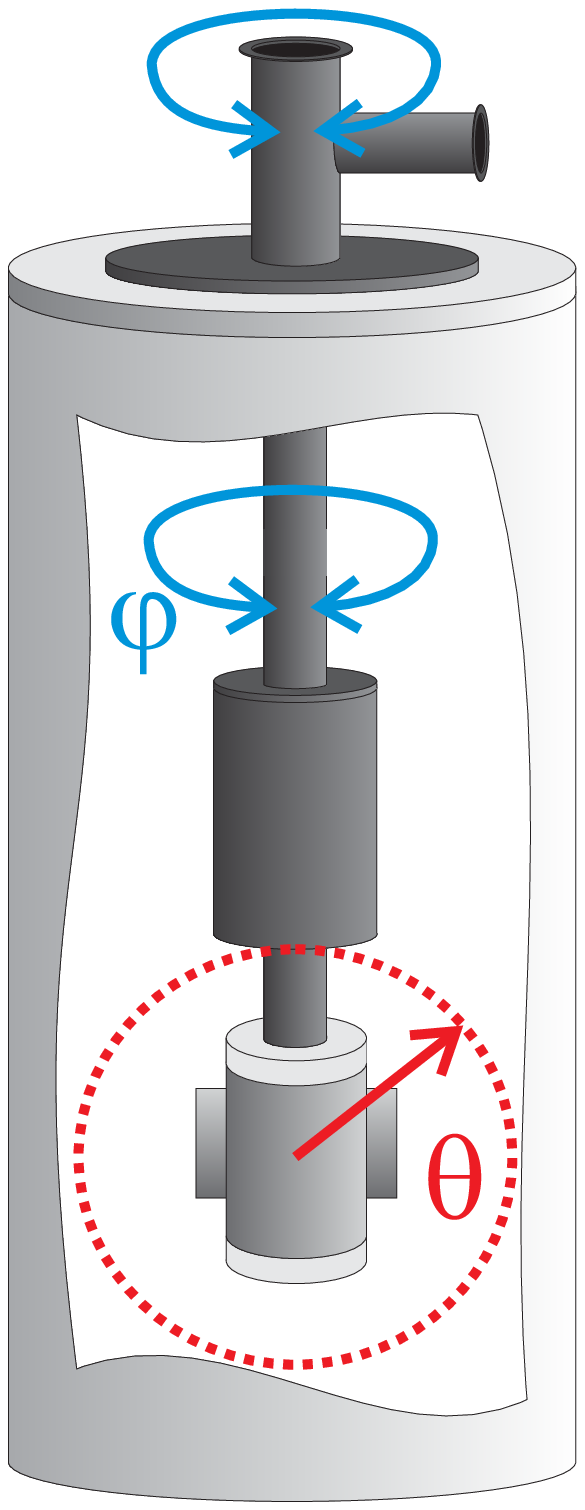}\label{fig:CryoRotationA}}\qquad
		\subfigure[]{\includegraphics[width=0.30\textwidth]{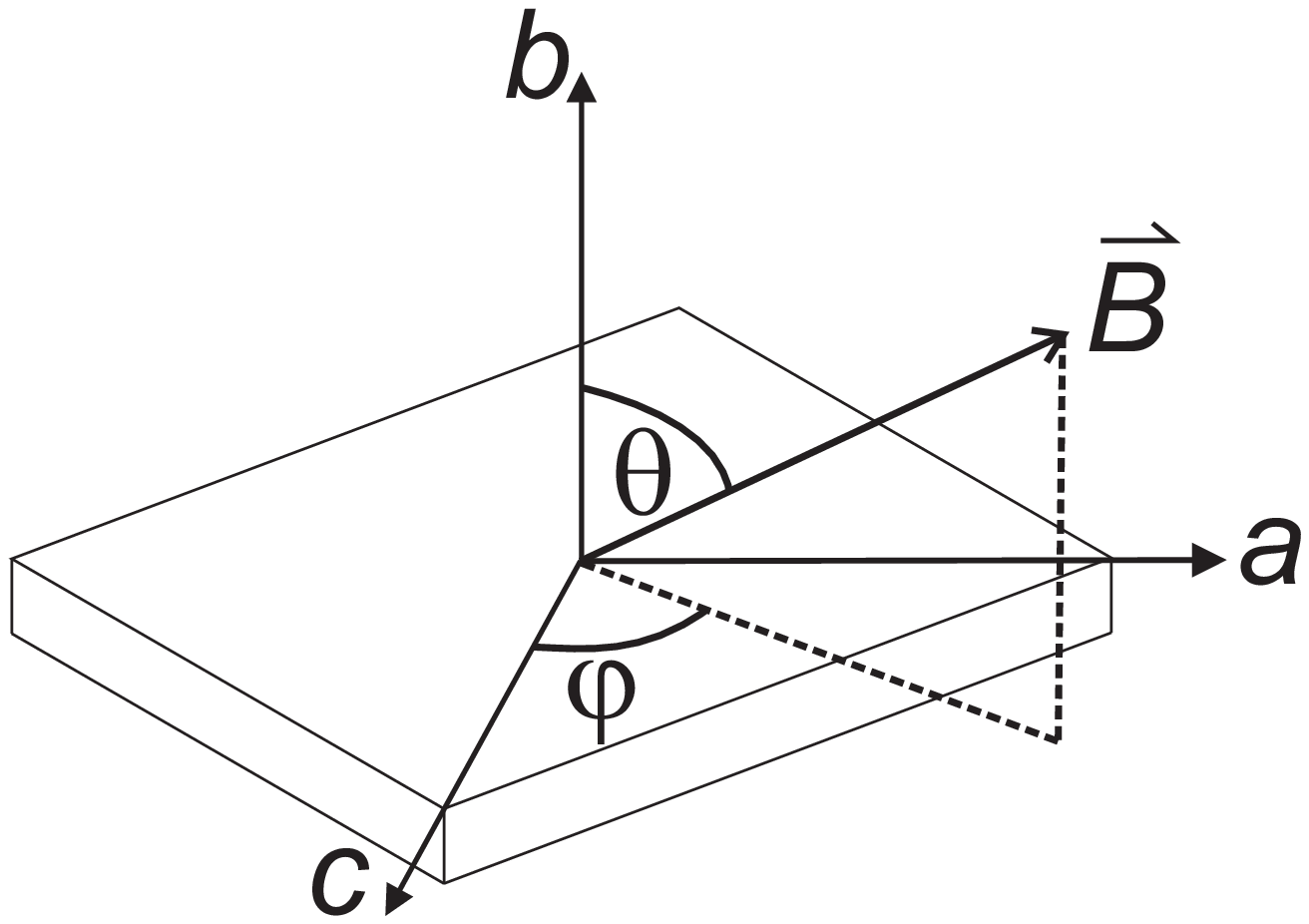}\label{fig:CryoRotationB}}
	\caption{(Color online) (a) Schematic drawing of the cryostat with the vector magnet system. The red lines indicate the field rotation in a vertical plane using the 2D vector magnet. The blue arrows demonstrate the manual rotation of the dilution-fridge insert inside the vector magnet, thus enabling a field rotation in all three dimensions. (b) Definition of the angles with respect to the magnetic field.}
	\label{fig:CryoRotation}
\end{figure}
For studying the angle-dependent magnetoresistance the crystals were placed in the center of a two-axes vector magnet with the maximal vertical and horizontal field components $\mu_0H_{\mathrm{v,max}} = 1.8$\,T and $\mu_0H_{\mathrm{h,max}} = 0.4$\,T, respectively, allowing rotation of the field in a vertical plane as depicted by the red lines in Fig.\,\ref{fig:CryoRotationA}.
Additionally, the cryostat with the vector magnet could be turned manually against the dilution unit with the sample stage, as shown by the blue arrows in Fig.\,\ref{fig:CryoRotationA}, thus changing the sample orientation with respect to the horizontal component of the field.
%In addition, the vector magnet could be turned manually against the dilution unit during operation shown by the blue arrows in Fig.\,\ref{fig:CryoRotationA}.
In the first measurement run the sample was oriented with its $b$-axis along the axis of the vertical coil.
In this geometry the inplane field orientation can be set with a high accuracy: $\Delta \theta < 0.1^{\circ}$ and $\Delta \varphi < 1^{\circ}$, where $\theta$ is the polar angle and $\varphi$ is the azimuthal angle inside the $ac$-plane, as shown in Fig.\,\ref{fig:CryoRotationB}.
However, the maximal inplane field in this case is limited to $B_\mathrm{h,max}=0.4$\,T. In two subsequent runs the crystal was aligned with its $c$- and $a$-axis parallel to the vertical field, respectively. In this geometry the error bar for the azimuthal orientation was slightly worse, $\Delta\varphi \leq 2^\circ$.

In total two different samples of $\kappa$-Cl were cooled down to lowest temperatures showing very similar behaviour to changes of temperature and perpendicular magnetic field. All the further studies on the phase diagram over several cooling cycles were done on only one of the samples. Therefore, only data from this sample are presented in this paper.

%\section{Results and Discussion}
\section{Temperature dependence of the interlayer resistance \label{sec:R(T)}}

\begin{figure}[tb]
	\centering
		\includegraphics[width=0.45\textwidth]{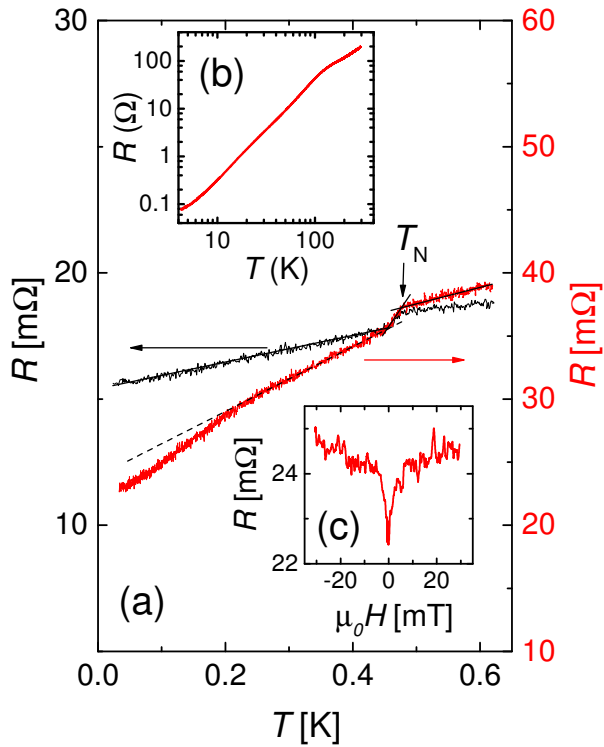}
	\caption{(Color online) (a) Interlayer resistance $R(T)$ of $\kappa$-(BETS)$_2$FeCl$_4$ below 0.6\,K for one sample in two different cooling runs with a measurement current of $I_s=1\,\mathrm{\mu A}$. The black curve (left-hand scale) shows the initial run; the red curve (right-hand scale) was obtained in the subsequent cooling cycle, after the crystal had been warmed up to room temperature and a part of it had been broken off. $T_{\mathrm{N}}$ is defined as the upper temperature of the resistance kink, as indicated by the arrow. (b) $R(T)$ curve for the same sample between 300\,K and 4\,K for the same cooling run as the red curve in (a). (c) Resistance of the same sample under a small magnetic field applied perpendicular to the layers, at $T = 22$\,mK.}
	\label{fig:R_T}
\end{figure}
Figure\,\ref{fig:R_T} shows the temperature dependence of the sample resistance. The overall behavior, see Fig.\,\ref{fig:R_T}(b) is very similar to the data reported earlier \cite{kobayashi_new_1993}.
The resistance monotonically decreases at cooling, showing a broad shallow hump between 100 and 200\,K but no peak characteristic of most $\kappa$-(BEDT-TTF)$_2$X salts\cite{ishi98,toyo07} and also observed in the sister compound $\kappa$-Br \cite{fujiwara_novel_2001,ojima_antiferromagnetic_1999,otsuka_organic_2001}.
The room- to low-temperature resistance ratio is among the highest obtained for organic charge transfer salts, reaching values $R(293\mathrm{K})/R(0.5\mathrm{K})\simeq 15,000$ for the samples studied.
In Fig.\,\ref{fig:R_T}(a) the black and red curves show the low-temperature $R(T)$ dependence of the same sample during two cooling cycles.
After the first cycle a part of the sample was broken off so that it had to be re-contacted. The difference in the absolute resistance value is caused by the reduced cross-section area of the sample in the second cycle. Both curves show a clear resistance drop (a "kink") by about 5\% at $T_{\mathrm{N}}=0.47$\,K, indicating the transition to the AFM state \cite{otsuka_organic_2001}.

Remarkably, a considerable linear temperature dependence of the resistance, $\mathrm{d}\ln R/\mathrm{d}T=0.15\,\mathrm{K}^{-1}$ and 0.37\,K$^{-1}$ for the black and red curves, respectively, is observed down to the AFM transition temperature. Moreover, it is even enhanced by about a factor of two below $T_{\mathrm{N}}$. The linear dependence, persisting down to lowest temperatures without saturation, is known for a number of other materials with strongly correlated electrons, including organic metals, heavy fermion compounds and
high-$T_c$ superconductors,
and interpreted as a signature of the non-Fermi-liquid behavior in the vicinity of a quantum critical point, see, e.g., Refs.\,\citenum{doiron-leyraud_correlation_2009,taillefer_scattering_2010,daou09}. It is, therefore, possible that the present salt is also close to a quantum phase transition. Indeed, on the one hand, the conduction system may be close to magnetic ordering triggered or assisted by the ordering of the adjacent $d$-electron system. On the other hand, the dimerized structure of the BETS layers and hence effectively half-filling of the conduction band can obviously lead to a Mott-insulating instability typical of many $\kappa$-type salts \cite{toyo07,ardavan2012}. It is, however, not clear whether electron correlations are sufficiently strong to cause a considerable Mott instability in the present case: the monotonic $R(T)$ dependence observed in the whole temperature range and particularly the very high resistance ratio observed would rather point to a good metallic character of the charge carriers. Further purposeful studies are needed to clarify this issue.

Despite the high crystal quality, evidenced, e.g., by the high resistance ratio, no clear manifestation of bulk superconductivity has been found in our measurements. In the first measurement run [black curve in Fig.\,\ref{fig:R_T}(b)] no sign of a SC transition has been observed. In the second run [red curve in Fig.\,\ref{fig:R_T}(a)] a weak downturn from the linear dependence can be seen below 0.21\,K (reproduced after thermal cycling between room and low temperatures). This downturn can be suppressed by a weak magnetic field below 10\,mT applied perpendicular to the layers, as shown in Fig.\,\ref{fig:R_T}(c). The onset temperature $T_0\approx0.21$\,K is consistent with the temperature range in which the a.c. susceptibitlity \cite{otsuka_organic_2001} and $\mu$SR \cite{pratt_sr_2003} anomalies suggesting a SC transition were observed. Therefore, it is likely associated with a formation of an inhomogeneous, filamentary SC state. If so, this is, to the best of our knowledge, the first manifestation of superconductivity in resistive properties of this compound. For example Otsuka \textit{et al.} \cite{otsuka_organic_2001} report $R(T)$ measurements down to 60\,mK without any sign of a SC transition. In our studies this feature was only seen in one sample.

It thus appears that superconductivity is extremely sensitive to minor crystal imperfections.
%The latter can be generated, for example, during the cooling process due to different thermal contractions of the electrical leads, contact pads, and the sample.
In the sister compound $\kappa$-Br the SC transition also shows a considerable dependence on crystal quality, as it follows from  comparing the data obtained by several groups \cite{fujiwara_novel_2001,fujiwara_indication_2002,konoike_magnetic-field-induced_2004,tanatar_thermal_2003,konoike_magnetic_2005,scha2014}. A possible reason for this is a nodal SC order parameter, which can be suppressed even by a small amount of nonmagnetic impurities. This scenario, extensively debated in relation to the $\kappa$-(BEDT-TTF)$_2$X salts, see, e.g., Refs. \citenum{toyo07,ardavan2012}, also looks plausible for the present $\kappa$-(BETS)$_2$X salts. Indeed, if, as noted above, the conducting system is close to an AFM quantum phase transition, this should favor a $d$-wave SC pairing mediated by AFM fluctuations. On the other hand, one should not disregard a possible role of internal strains. A very strong dependence of superconductivity on pressure is a general feature of organic superconductors \cite{toyo07}. Taking into account the very low $T_c$, it is not excluded that strains appearing at cooling due to different thermal contraction of the sample and the electrical contacts (graphite paste) have a strong impact on superconductivity in this material. To check whether this is the case, it would be interesting to perform comparative studies of one and the same crystal using different techniques, e.g., with and without electrical contacts.

\section{Magnetoresistance and Shubnikov-de Haas effect \label{sec:SdH}}

\subsection{Manifestations of the AFM state in magnetoresistance}

\begin{figure}[t]
	\centering
		\includegraphics[width=0.45\textwidth]{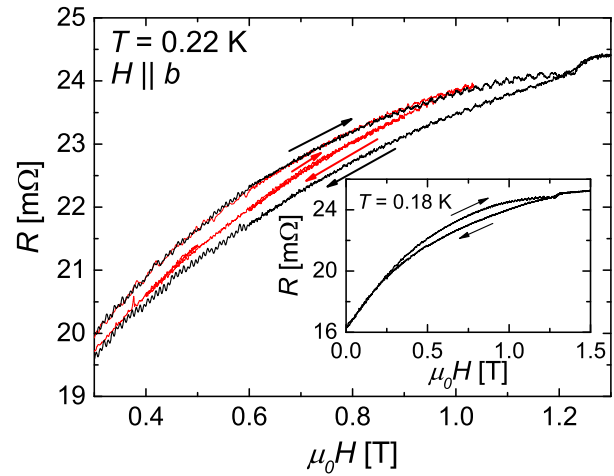}
	\caption{(Color online) Interlayer resistance as a function of magnetic field perpendicular to conducting layers near the AFM -- PM transition. Sweep from zero field until above the transition and back (black curve) and a sequence of up- and down-sweeps between intermediate field values (red curve). Before recording the red curve the field was raised from 0 to 1.03\,T; then the data was taken at the following sweep sequence: $1.03\,\mathrm{T}\rightarrow 0.8\,\mathrm{T}\rightarrow 0.9\,\mathrm{T}\rightarrow 0.6\,\mathrm{T}\rightarrow 0.8\,\mathrm{T}\rightarrow 0.4\,\mathrm{T}\rightarrow 0.5\,\mathrm{T}\rightarrow 0\,\mathrm{T}$. The inset shows the full field range with the complete hysteresis loop for the lowest measured temperature.}
	\label{fig:R_B-perp_hys}
\end{figure}
In Fig.\,\ref{fig:R_B-perp_hys} the magnetic field dependence of the interlayer resistance for the field direction perpendicular to the layers is shown for $T=0.18\,\mathrm{K}$. At $\mu_0 H_k=1.3$\,T a resistance step similar to the kink feature in the temperature sweep [Fig.\,\ref{fig:R_T}(a)] is observed. Because of the similarity to the feature in $R(T)$, we suggest that this is the transition from the AFM to the paramagnetic (PM) state. This guess is substantiated by the observation of Shubnikov-de Haas (SdH) oscillations which exist below and abruptly vanish above $H_k$, as will be presented in the next Section.

The field-dependent resistance in the low-field AFM state is characterized by a strong hysteresis, implying the presence of a domain structure. The hysteresis loop is fully reproducible by sweeping the field up to $H_k$ and back to zero. However, if the sweep direction is inverted at a field within the hysteresis range, the resistance shows a reversible behavior, continuously changing between the upper and lower branches of the full hysteresis loop, as shown by the red line in Fig. \ref{fig:R_B-perp_hys}. The exact trace $R(H)$ is thereby only determined by the value of the highest field applied in the sequence.

A hysteresis in the interlayer magnetoresistance has also been reported for the $\kappa$-Br salt \cite{konoike_anomalous_2007}.  As opposed to our case, in that salt a clear difference between the initial up-sweep and the following down-sweep of the field was observed all the way from $H_k$ down to the SC transition at which the resistance dropped to zero. Furthermore, in consecutive field cycles the resistance traces fully reproduced the first down-sweep, showing that the "memory" of the high-field state was preserved even in zero field. In our case the memory is obviously lost at fields below $\approx 0.25$\,T, where the hysteresis vanishes, see inset in Fig.\,\ref{fig:R_B-perp_hys}.

For the $\kappa$-Br salt it was proposed that the hysteresis originates from additional scattering on domain walls, which may appear upon entering the AFM state from the high-field PM state with saturated Fe$^{3+}$ spins \cite{konoike_anomalous_2007}. On the other hand, our data in Fig.\,\ref{fig:R_B-perp_hys} shows that crossing $H_k$ is not necessary for producing the hysteresis: the domains apparently arise already at cycling the field within the spin-canted AFM phase. Despite the above-mentioned small differences in the hysteresis behavior, it is most likely that the origin of the hysteresis is common for the two salts while exact details are determined by the crystal imperfections and particular domain distribution in the sample.

We note that the resistance is lower in the down-sweep than in the up-sweep both in $\kappa$-Br \cite{konoike_anomalous_2007} and in $\kappa$-Cl. At first glance, this contradicts the suggested enhancement of scattering in the down-sweeps. However, taking into account a very high anisotropy of the present compounds, it is possible that defects like domain walls provide an additional, incoherent channel to the interlayer conductivity in parallel to the conventional coherent one \cite{kart09b,anal06}. Thus, such defects can lead to a decrease of the interlayer resistivity, as observed in the experiment.

\subsection{SdH oscillations in the AFM state  \label{sec:SdHAFM}}

\begin{figure}[tb]
	\centering
		\includegraphics[width=0.45\textwidth]{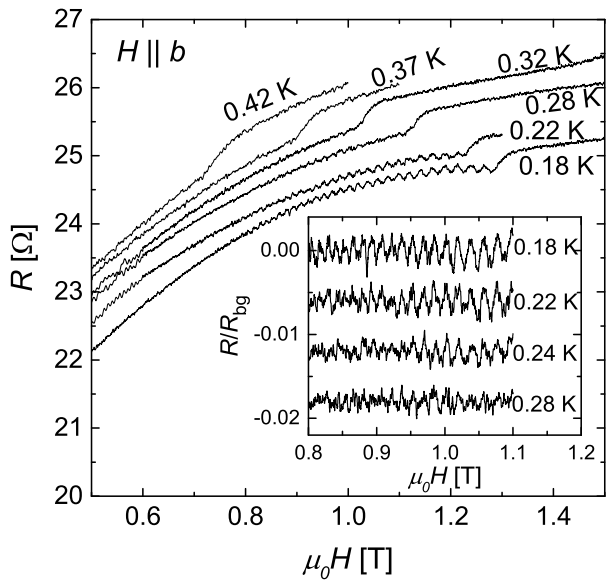}
	\caption{Magnetic field sweeps at different temperatures with the field applied perpendicular to the conducting layers. The inset shows the oscillatory patterns used for determination of the cyclotron mass; the curves are obtained by dividing the as-measured $R(H)$ data by the nonoscillating background resistance $R_{\mathrm{bg}}(H)$.}
	\label{fig:R_B-perp_mc}
\end{figure}
Thanks to the very high crystal quality, we were able to observe slow SdH oscillations starting from below 1\,T. Examples of the oscillatory resistance recorded at several temperatures are shown in Fig.\,\ref{fig:R_B-perp_mc}. One can see that the oscillations only exist below $H_k$ and thus are an inherent feature of the AFM state. The oscillation frequency,  $F = 58$\,T, corresponds to a small Fermi surface cross section occupying 1.4\% of the first Brillouin zone area.

\begin{figure}[tb]
	\centering
		\includegraphics[width=0.45\textwidth]{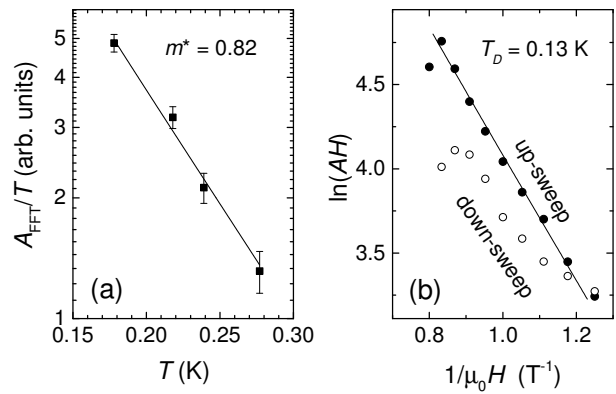}
	\caption{(a) Temperature dependence of the SdH amplitude (symbols) and the fit by the Lifshitz-Kosevich formula \cite{ShoenbergBook1984} (line) yielding the normalized cyclotron mass $m^* = 0.82$. The amplitude $A_{\mathrm{FFT}}$ was taken as the peak height in the fast Fourier transform of the oscillatory pattern shown in the inset in Fig. \ref{fig:R_B-perp_mc}. (b) "Dingle" plot of the oscillation amplitude recorded at $T=0.18$\,K in the up- (filled circles) and down- (open circles) sweeps of the field. The line is the fit of the up-sweep data (see text) with the Dingle temperature $T_D = 0.13$\,K.}
	\label{fig:mass&dingle}
\end{figure}

With increasing temperature the transition field $H_k$ decreases, as we can see in Fig.\,\ref{fig:R_B-perp_mc}, so the window for the observation of the SdH oscillations becomes more narrow. As a result, the determination of the effective cyclotron mass, from the temperature dependence of the oscillation amplitude \cite{ShoenbergBook1984}, is only possible in a very restricted field and temperature range: 0.8 to 1.1\,T and 0.18 to 0.27\,K, respectively. The corresponding plot based on the data presented in the inset in Fig.\,\ref{fig:R_B-perp_mc} is shown in Fig.\,\ref{fig:mass&dingle}(a). Fitting the data by the standard Lishitz-Kosevich formula, we obtain a cyclotron mass value of $m^* = 0.8\pm 0.1$, expressed in units of the free electron mass. This mass is very high for such a small Fermi surface. It implies that many-body interactions and correlation effects are important in this system, which is in line with the linear temperature dependence at low temperatures discussed in the previous section.

The field dependence of the oscillation amplitude at $T=0.18$\,K is presented in Fig.\,\ref{fig:mass&dingle}(b) in the form of a Dingle plot for a quasi-2D metal \cite{ShoenbergBook1984,grig03,kart04}. The up-sweep data, except one highest-field point \cite{comment_Dingle}, shows a conventional behavior which can be fitted by a straight line with the slope
$K \approx -am^*(T+T_D)/\mu_0H$, where $a = 14.69$\,T/K and $T_D = \hbar/2\pi k_B\tau$ is known as the Dingle temperature. By fitting the data, we obtain a very low Dingle temperature $T_D = 0.14$\,K. This corresponds to a long scattering time, $\tau = \hbar/2\pi k_B T = 8.6$\,ps, another evidence of a very high crystal quality. On the down-sweep, the amplitude is considerably lower and clearly violates the conventional behavior. This enhanced damping of the oscillations in the down-sweep is obviously caused by the same additional, field-dependent scattering that causes the hysteresis of the nonoscillating magnetoresistance presented above.

SdH oscillations very similar to those presented here have also been observed in the AFM state of the $\kappa$-Br salt \cite{konoike_fermi_2005,konoike_anomalous_2007}.  The magnetic ordering in $\kappa$-Br is more robust: in the field perpendicular to the layers it survives up to 5.5\,T, see, e.g. ref. \citenum{konoike_fermi_2005}. That is 4 times higher than for the present compound. Nevertheless, the main oscillation parameters, $F=62$\,T and $m^* = 1.1$, are close to what we find for $\kappa$-Cl.

The fact that the present oscillations only exist in the AFM state implies that the Fermi surface is reconstructed by the magnetic superstructure. This is also corroborated by the absence of small pockets, which could give rise to the experimentally obtained low SdH frequency, in the calculated original, nonmagnetic Fermi surface \cite{kobayashi_new_1996}. Konoike \textit{et al.} \cite{konoike_fermi_2005} have proposed a plausible reconstruction based on the theoretically predicted \cite{mori_estimation_2002} superstructure wave vector $\mathbf{Q}_{\mathrm{AFM}} = (0,0,\pi/c)$. They, indeed, have obtained small Fermi pockets consistent with the observed SdH frequency. The reconstructed multiply connected Fermi surface contains a number of other, bigger closed pockets, which should, in principle, also contribute to SdH oscillations. However, taking into account that the relevant cyclotron masses are expected to be higher, the amplitudes should be much stronger suppressed in the low-field range corresponding to the AFM state. We note, that a very weak oscillatory component with a frequency $\approx 3$ times higher than the fundamental one has been reported by Konoike et al.\cite{konoike_anomalous_2007}. Further detailed studies on high-quality samples are needed to verify whether it indeed originates from another part of the reconstructed Fermi surface in the AFM state and is not just a strong 3rd harmonic of the fundamental frequency caused by high two-dimensionality of the charge carriers \cite{kart04,harr96c}.

\section{Phase diagram}
The pronounced steplike anomaly in the interlayer resistance at the AFM -- PM transition (see Figs.\,\ref{fig:R_T} and \ref{fig:R_B-perp_hys}) can be utilized for establishing the magnetic phase diagram. For each of the three crystal axes the transition was studied in isothermal magnetic field sweeps at different temperatures starting with the lowest temperature up to $0.9\,T_{\mathrm{N}}$. In addition, the phase boundary was studied by doing temperatures sweeps at different magnetic fields.

\subsection{Magnetic field along the easy axis}

\begin{figure}[tb]
	\centering
		\includegraphics[width=0.45\textwidth]{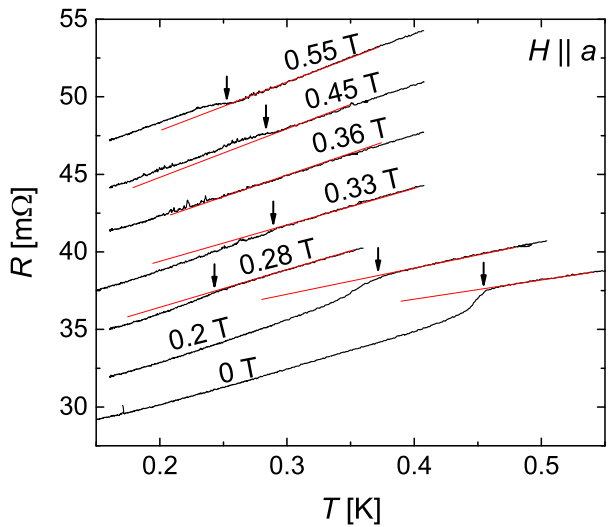}
	\caption{(Color online) Temperature sweeps at different magnetic fields parallel to the $a$-axis. All the curves, except one at zero field, have been shifted for better visibility. The red lines are guides to the eye, showing the slope above the transition. The arrows mark the points taken as transition points.}
	\label{fig:R_B-a-axis_T}
\end{figure}

According to the a.c.-susceptibility measurements \cite{otsuka_organic_2001}, the easy axis of magnetization is along the crystallographic $a$-axis.
%The most interesting direction is the easy axis of the magnetization, which was determined by ac susceptibility measurements to lie along the inplane $a$-axis. \cite{otsuka_organic_2001}
In Fig.\,\ref{fig:R_B-a-axis_T} temperature sweeps at different values of magnetic field aligned in this direction are presented, showing two notable features. Firstly, the transition temperature rapidly decreases at increasing the field till 0.28\,T. However, the monotonic decrease is interrupted in the field interval between 0.28 and 0.33\,T where the transition shifts up by $\approx 40$\,mK. Secondly, the resistance anomaly decreases in size for increasing fields and becomes unresolvable between 0.33\,T and 0.45\,T. At 0.45\,T it reappears with an inverted sign.

\begin{figure}[b]
	\centering
		\includegraphics[width=0.45\textwidth]{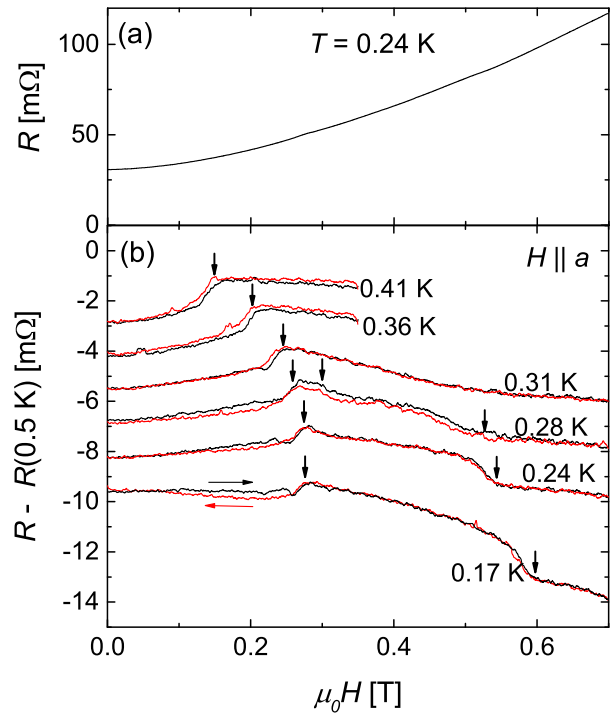}
	\caption{(Color online) (a) Example of a magnetic field sweep at $H\|a$. The AFM--PM transition is hardly visible. (b) Field-sweep curves at $H\|a$ after subtraction of the curve taken at $T=0.5$\,K $> T_{\mathrm{N}}$. Up- and down-sweeps are colored black and red, respectively, as marked by the horizontal arrows. The vertical arrows mark the transition points.}
	\label{fig:R_B-a-axis_B}
\end{figure}

For this direction of magnetic field the magnetoresistance is quite high [see Fig.\,\ref{fig:R_B-a-axis_B}(a)], showing an approximately $H$-squared dependence in the present field range. The anomaly due to the phase transition is hard to resolve, mainly because of the strong monotonic background. However, after subtracting the field-dependent signal recorded at $T=0.5$\,K, i.e. immediately above the zero-field transition temperature, clear features are observed, as shown in Fig.\,\ref{fig:R_B-a-axis_B}(b).

Near the N\'{e}el temperature, for example at 0.41\,K, the field-induced AFM -- PM transition is manifested by a distinct increase of the resistance. This is of course consistent with the character of the resistance change in the zero- and low-field temperature sweeps through the transition.
There seems to be a small hysteresis between the up- and down-sweeps at the transition step for temperatures near $T_{\mathrm{N}}$. But it is too weak to judge whether it has physical reasons. Away from the transition step no hysteresis was detected in the $R(H)$ sweeps. At low temperatures we see a second significant step also marked by an arrow in Fig.\,\ref{fig:R_B-a-axis_B}(b), e.g. at $\mu_0 H=0.59$\,T in the 0.18\,K curve.

\begin{figure}[tb]
	\centering
		\includegraphics[width=0.5\textwidth]{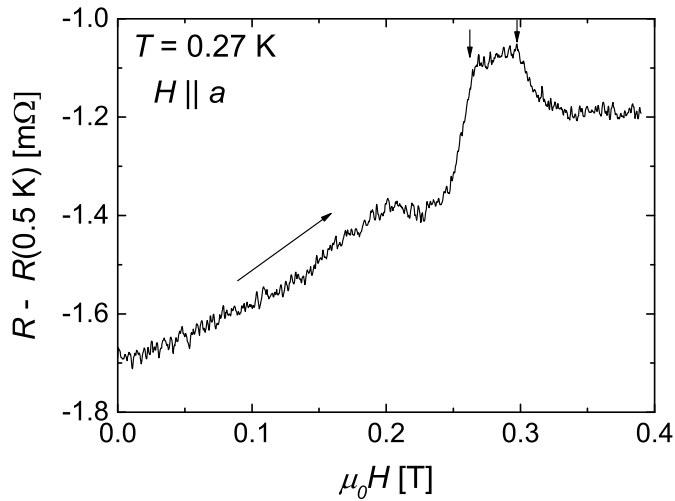}
	\caption{A field sweep, up to 0.4\,T, at a precise field orientation, $H\|a$. There are two transitions close to each other, as indicated by the vertical arrows. The second one is the reentrant transition into the spin-flopped AFM$_1$ state, see Fig.\,\ref{fig:PD_a-axis}.}
	\label{fig:R_B-a-axis_reentrant}
\end{figure}
In the narrow temperature range $0.25\,\mathrm{K}\leq T<0.31\,\mathrm{K}$ even a third feature can be resolved [see the curve for $T=0.28$\,K in Fig.\,\ref{fig:R_B-a-axis_B}(b)].
This additional feature is best pronounced when the field is precisely aligned along the $a$-axis. An example of a sweep in the exactly oriented field, up to 0.4\,T is shown in Fig.\,\ref{fig:R_B-a-axis_reentrant}. Here the resistance displays a step up at $\mu_0 H=0.26$\,T and a step down at $\mu_0 H=0.30$\,T. The higher-field sweeps presented in Fig.\,\ref{fig:R_B-a-axis_B}(b) were carried out in the configuration, in which the alignment was less precise, as explained in Section \ref{sec:experimental}. This, most likely, is the reason why the reentrant transition is weaker pronounced in the 0.28\,K curve in this Figure.
\begin{figure}[tb]
	\centering
		\includegraphics[width=0.5\textwidth]{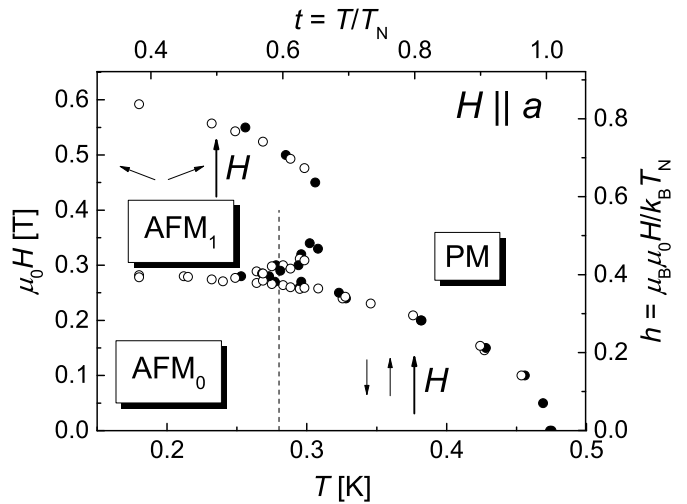}
	\caption{\textit{H-T} phase diagram for $H\|$a-axis. Empty and filled symbols correspond to \textit{H}- and \textit{T}-sweeps, respectively. Arrows show schematically the arrangement of the staggered Fe$^{3+}$ electron spins with respect to the external field in the low-field AFM$_0$ and high-field AFM$_1$ antiferromagnetic states, respectively. The vertical dashed line shows the section of the phase diagram by the field sweep in Fig.\,\ref{fig:R_B-a-axis_reentrant}.}
	\label{fig:PD_a-axis}
\end{figure}

The resulting phase diagram for $H\|a$-axis is shown in Fig.\,\ref{fig:PD_a-axis}. It presents a textbook example for an uniaxial antiferromagnet \cite{LandauLifshitz1984,BlundellBook2001} with a bicritical point at $T^{\ast} = 0.25\,\mathrm{K} \approx 0.56 T_{\mathrm{N}}$ and $\mu_0 H^{\ast} = 0.28\,\mathrm{T}$. The high-field, low-temperature phase AFM$_1$ is most likely a spin-flopped AFM state \cite{BlundellBook2001,LandauLifshitz1984,fisher_scaling_1975,fisher_spin_1974,landau_phase_1978}. This means that in the low-temperature regime a spin-flop transition takes place at about 0.28\,T: the spins turn by $\sim 90^\circ$ and form an AFM order with the staggered magnetization direction perpendicular to the external field and with a small ``ferromagnetic'' component along the field, as schematically illustrated in Fig.\,\ref{fig:PD_a-axis}. According to theory, the spin-flop is a first order phase transition, as the total magnetization changes discontinuously at the transition. Therefore, some kind of hysteresis at the spin-flop transition would be expected. However, in our experiments we could not detect a clear hysteresis at the transition.

The AFM$_1$ phase is suppressed in a second order phase transition at a field, which, at the lowest temperatures in our experiment, is approximately double the value of the spin-flop field. When going to higher temperatures, this upper transition moves to lower fields while the spin-flop field stays approximately constant until the bicritical point $T^{\ast}$. The third feature in the fields sweeps made immediately above $T^{\ast}$ is obviously associated with the reentrant transition from the PM state to the spin-flopped AFM$_1$ state.
For temperatures above 0.32\,K the AFM$_1$ phase vanishes completely.

In Fig.\,\ref{fig:PD_a-axis}, there is a gap in the data set delineating the highest-temperature part of the AFM$_1$/PM phase boundary. The corresponding field interval, between 0.33 and 0.45\,T, is exactly where the resistance becomes practically insensitive to the transition. As mentioned at the beginning of this Section, this happens because the kink feature in the $R(T)$ dependence is changing its sign: at lower fields the resistance decreases upon entering the AFM state, whereas a weak increase is detected at $\mu_0H\geq 0.45$\,T, see Fig.\,\ref{fig:R_B-a-axis_T}. An explanation of this behavior should obviously lie in the coupling of the charge transport to the spin system. The resistance decrease observed at low fields seems to be a natural consequence of a reduced spin-dependent scattering in a magnetically ordered state. The effect of increasing field is to align localized spins in one direction, which leads to a decrease in the spin-dependent scattering even in the PM state and thereby to a decreasing difference between the resistances in the PM and AFM states. It could then happen that the resistance in the AFM state becomes even somewhat higher in the AFM state because of disorder in the magnetic structure. This all seems to be consistent with the data in Fig.\,\ref{fig:R_B-a-axis_T}. However, at present we have no convincing explanation why the inversion of the kink feature is particularly pronounced for the field along the easy magnetization axis $a$ and not observed, for example, at $H \| b$: in the latter case the resistance always drops on entering the AFM state, as one can see in Fig.\,\ref{fig:R_B-perp_mc}.
%However, it is even very sensitive to an exact orientation of $\|a$: For a magnetic field tilted by $5^\circ$ from $a$-axis in the $ac$-plane the behavior is also different. Under these conditions the transition step in the temperature sweeps is observed at all fields up to $B_k$ and never changed its sign. This makes us sure that the missing points are just an artifact of the interlayer resistance measurement. The phase diagram for this field orientation looks qualitatively the same as in Fig.\,\ref{fig:PD_a-axis}, with the only difference that the phase line around the bicritical point is smeared out meaning that almost no decrease in the transition temperature at the bicritical field is observed.
%On the other hand, such a scenario is expected to be, at least qualitatively, independent of the direction of the magnetic field. By contrast, the data

The $H$--$T$ phase diagram in Fig.\,\ref{fig:PD_a-axis} shows notable differences from that reported for $\kappa$-Br. For the latter salt no clear evidence of a spin-flop transition has been found,
see, e.g., Ref.\,\citenum{konoike_magnetic-field-induced_2004}. In principle, in resistive properties the spin-fop transition could be hidden inside the zero- or low-resistance SC state. In magnetic torque measurements \cite{konoike_magnetic_2005} two distinct features observed at 1.7\,T and 1.9\,T, for $H \| a$, were suggested to originate from the spin-flop and AFM -- PM transitions, respectively. If this is true, the relative field range of the spin-flopped phase is much more narrow for $\kappa$-Br than for $\kappa$-Cl. This is likely a consequence of a higher inplane anisotropy of the $\kappa$-Br salt, as will be demonstrated below.

For a more quantitative comparison between the phase diagrams of the two sister compounds, which will be done in the following Section, it is convenient to introduce reduced temperature and magnetic field: $t=T/T_{\mathrm{N}}$, $h=\mu_0\mu_B H \,/\, k_B T_{\mathrm{N}}$. In Fig.\,\ref{fig:PD_a-axis} these reduced coordinates are given on the top and right-hand axes, respectively. Now one can estimate that in our salt the spin-flop transition occurs at $h_{\mathrm{sf}} \approx 0.4$, which is close to the critical value $h_k^a$ for the AFM -- PM transition in $\kappa$-Br at $t=0.5$ \cite{konoike_magnetic-field-induced_2004}. At the same time the transition to the PM state in our salt has approximately double this value, $h_k^a=0.78$.

\subsection{Other field orientations}

\begin{figure}[tb]
	\centering
		\includegraphics[width=0.45\textwidth]{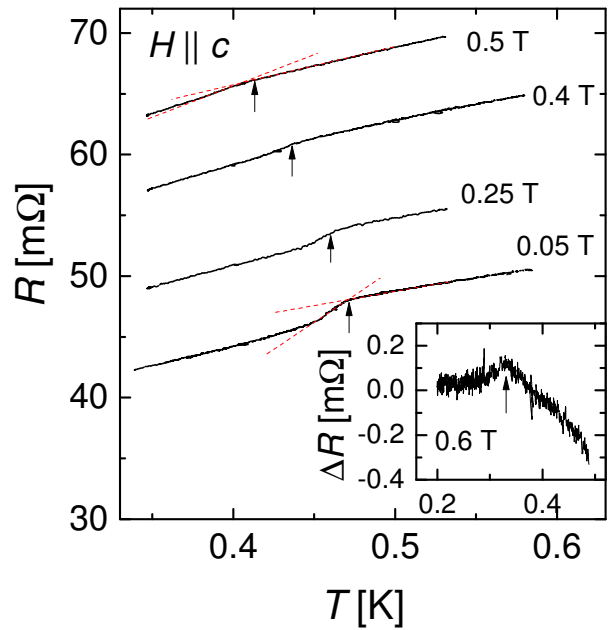}
	\caption{Examples of $T$-dependent resistance traces at different fields directed along the inplane $c$-axis. The arrows point to the crossing points of the linear extrapolations from the normal-state $R(T)$ and the resistance kink (shown by dashed red lines for the 0.5 and 0.05\,T curves), defined as transition points. Inset: Data recorded at $\mu_0H=0.6$\,T after subtracting a cubic fit.}
	\label{fig:R_B-c-axis_B-sweeps}
\end{figure}
Next we consider the influence of a magnetic field applied parallel to the crystallographic $b$- and $c$-axes. Like in the previous case, the AFM state is suppressed by a sufficiently high field, which is reflected in the kink feature in the temperature- and field-dependent resistance. However, since now the field is perpendicular to the easy magnetization axis, no spin flop takes place. At $H\|b$ the AFM -- PM transition is clearly seen in raw temperature and field sweeps, as demonstrated, for example, in Fig.\,\ref{fig:R_B-perp_mc} for $H$-sweeps. At $H\|c$, the kink is well pronounced at low fields; examples are shown in Fig.\,\ref{fig:R_B-c-axis_B-sweeps}. With increasing the field the kink becomes weaker. Therefore, background subtraction has been applied in order to determine the transition point at fields $\geq 0.5$\,T, see e.g., inset in Fig.\,\ref{fig:R_B-c-axis_B-sweeps}. However, unlike for the $B\|a$ orientation, the kink feature does not change its sign.

\begin{figure}[tb]
	\centering
		\includegraphics[width=0.5\textwidth]{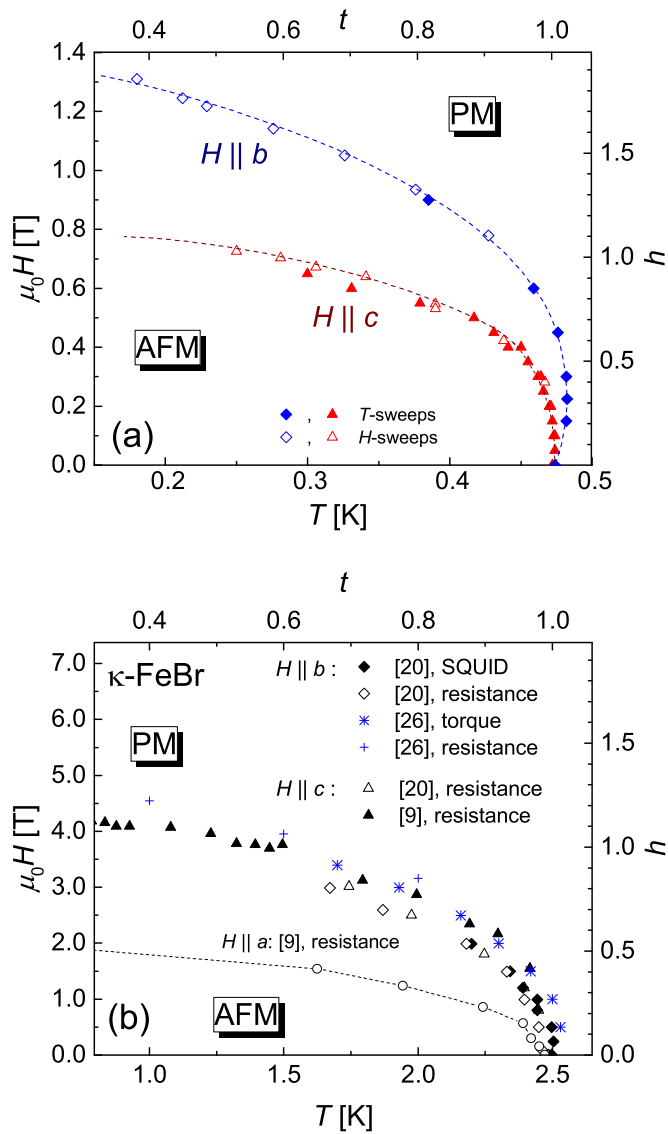}
	\caption{(Color online) (a) \textit{H--T} phase diagram for $H\|b$ (blue) and $H\|c$-axis (red). Filled and empty correspond to \textit{T}- and \textit{H}-sweeps, respectively. (b) \textit{H--T} phase diagram of $\kappa$-Br for three crystal axes. The data are taken from other works \cite{konoike_magnetic-field-induced_2004,fujiwara_novel_2001,scha2014}. Black open circles are for $B\|a$ and red triangles for $B\|c$. Blue diamonds, stars and crosses correspond to $B\|b$.}
	\label{fig:PD_b&c-axes}
\end{figure}
The transition points determined for $H\|c$ and for $H\|b$ are shown in Fig.\,\ref{fig:PD_b&c-axes}(a) by triangles and diamonds, respectively; open symbols were obtained from $H$-sweeps and filled symbols from $T$-sweeps. Taken together with those plotted in Fig.\,\ref{fig:PD_a-axis}, the data reveal a considerable anisotropy of the magnetic field effect in all three directions. At $T=0.5T_{\mathrm{N}}$ the ratio between the corresponding critical field values is: $H_k^a:H_k^b:H_k^c \approx 1:2.2:1.3$.
%$H_k^a:H_k^b:H_k^c = 0.56\,\mathrm{T}:1.22\,\mathrm{T}:0.74\,\mathrm{T} = 1:2.18:1.32$
%Comparing these results with the data on the $\kappa$-Br salt, shown in Fig.\,\ref{fig:PD_b&c-axes}(b), one can notice a few interesting points.
The obtained results can be compared with the reported data on the $\kappa$-Br salt \cite{konoike_magnetic-field-induced_2004,fujiwara_novel_2001,scha2014}, which are shown in Fig.\,\ref{fig:PD_b&c-axes}(b). To facilitate the comparison, axes with the normalized coordinates $t$ and $h$ are added to both graphs in Fig.\,\ref{fig:PD_b&c-axes}. One can immediately see that in these coordinates the phase lines for $H\|c$ are very similar for the two compounds, whereas for the other directions they are quite different.

As mentioned above, for $H\|a$ the normalized critical field of the AFM -- PM transition in the $\kappa$-Cl salt is considerably higher than in $\kappa$-Br at low temperatures. Thus, the difference between the critical fields along the two inplane principal axes, $h_k^a$ and $h_k^c$, is smaller in $\kappa$-Cl. This is most likely the reason for the existence of the spin-flop transition in the present salt, by contrast to $\kappa$-Br.

The weaker inplane anisotropy of $\kappa$-Cl seems to be in line with the theoretically predicted larger role of $\pi$-$d$ coupling in the AFM order \cite{mori_estimation_2002}: the dominant $\pi$-$d$ interaction ($J_6$ in notations of Ref.\,\citenum{mori_estimation_2002}) provides coupling in both the $a$- and the $c$-directions, whereas the direct $d$-$d$ coupling is obviously much stronger along the $a$-axis.

One may expect that a relatively large role of $\pi$-$d$ interactions in $\kappa$-Cl should also result in a stronger coupling in the third, interlayer direction. However, a comparison of the anisotropies in the $b$-$c$ plane apparently contradicts this expectation. While the critical fields along the $b$- and $c$-axes are practically the same for the $\kappa$-Br salt, in $\kappa$-Cl the $b$-axis critical field is $\approx 70\%$ higher than $h_k^c$, which suggests weaker interlayer magnetic correlations.

Furthermore, a careful inspection of the transition temperature at low fields for $B\|b$ reveals its slight increase at increasing field from 0 to $0.2$\,T, see Fig.\,\ref{fig:PD_b&c-axes}(a). The effect is small, $\Delta t/t\approx 8$\,mK or $1.7\%$ of $T_{\mathrm{N}}$, but definitely exceeds the experimental error bar. Only for $\mu_0 H>0.3\,\mathrm{T}$ the phase boundary line acquires the conventional negative slope. A similar effect has been observed in the quasi-2D antiferromagnet [Cu(HF)$_2$(pyz)$_2$]BF$_4$ and attributed to a suppression of phase fluctuations by a magnetic field \cite{sengupta_nonmonotonic_2009}. In zero field, the transition temperature of a quasi-2D antiferromagnet is diminished compared to the value one would obtain from mean field calculations. The reason for this are phase fluctuations, which suppress long range ordering. The fluctuations are reduced, when a magnetic field is applied and therefore $T_{\mathrm{N}}$ increases. At higher fields the suppression of the AFM state due to the increasing Zeeman energy becomes the dominant mechanism, leading to a restoration of the conventional negative slope of the phase line \cite{sengupta_nonmonotonic_2009}.

Comparing the $\kappa$-Cl and $\kappa$-Br salts, in the latter compound the $h(t)$ dependence is very steep near $t=1$, both for $H\|b$ and for $H\|c$. However, no increase of the transition temperature at low fields can be inferred from the data in Fig.\,\ref{fig:PD_b&c-axes}(b). Thus, we conclude that the fluctuations are relatively weak and the interlayer coupling is stronger in the $\kappa$-Br salt.

\section{Summary}

Aimed at a better understanding of the interplay of the magnetic and conducting subsystems in $\kappa$-(BETS)$_2$FeCl$_4$, we have performed detailed studies of its low-temperature interlayer resistance. A non-saturating, linear $R(T)$ dependence has been observed below 1\,K, which might be a signature of strong correlations in the conduction system in the vicinity of a magnetic quantum critical point. The samples investigated did not show a bulk SC transition. However, in one experiment a small downturn of the $R(T)$ curve below 0.2\,K, originating most likely from superconductivity arising in a tiny sample fraction, has been found. Thus, the resistive experiment corroborates the earlier reports \cite{otsuka_organic_2001,pratt_sr_2003} on a SC transition in this compound, pointing, however, that superconductivity is very weak and sample dependent.

Slow SdH oscillations found at fields below 1.3\,T, within the AFM state, suggest a reconstruction of the FS due to the magnetic ordering, similarly to the sister compound $\kappa$-(BETS)$_2$FeBr$_4$ \cite{konoike_fermi_2005}. The FS orbit responsible for the oscillations occupies only $1.4\%$ of the original (unreconstructed) first Brillouin zone but is characterized by a rather heavy cyclotron mass, $\approx 0.8$ of the free electron mass. This is regarded as another indication of strong correlations in the conducting system.

The kink features in $R(T)_{H=\mathrm{const.}}$ and $R(H)_{T=\mathrm{const.}}$ curves associated with the AFM transition make it possible to determine the $H$-$T$ phase diagram. This has been done for three field directions corresponding, respectively, to the principal magnetization axes, which coincide with the main crystallographic axes in the present compound.

For $H\|a$-axis, the easy axis of the Fe$^{3+}$ spin system, a clear evidence for a spin-flop transition was found. The field, at which the spin-flopped phase is broken, is about twice as high as the spin-flop field. For the field applied along the two hard axes ($b$- and $c$-axes) the phase diagram looks simpler, with only one AFM phase.
%The AFM state is most robust for magnetic fields along the $b$-axis. For the direction along the $c$-axis the transition fields are almost a factor of two smaller.
A comparison of these phase diagrams to those obtained for the $\kappa$-Br salt reveals a considerably lower inplane anisotropy of critical field. This explains the presence of the spin-flopped phase in an extended part of the phase diagram, by contrast to $\kappa$-Br. The revealed weaker inplane anisotropy seems to support the prediction of an enhanced relative contribution of $\pi$-$d$ interactions in setting the AFM order in $\kappa$-Cl \cite{mori_estimation_2002}. However, it is not quite clear at present how to reconcile this prediction with the considerably stronger in- to out-of-plane anisotropy, as compared to the $\kappa$-Br salt.

\begin{acknowledgments}
This work was supported by the German Research Foundation (DFG) via the grant KA 1652/4-1.
\end{acknowledgments}


\begin{thebibliography}{40}
\bibitem [{com()}]{comment_Dingle}%
  \BibitemOpen
  \href@noop {} {}\bibinfo {note} {The decrease of the SdH amplitude at
  $1/\mu_0H = 0.8$\,T$^{-1}$ is caused by proximity to the critical field $H_k$
  at which the oscillations vanish.}\BibitemShut {Stop}%
\end{thebibliography}
\end{document}